\documentclass[twocolumn,asmfonts,amssymb,amsmath,fleqn,floatfix]{revtex4}
\usepackage[english]{babel}
\usepackage[T1]{fontenc}
\usepackage{fixmath}
\usepackage[Symbol]{upgreek}
\usepackage[upmu]{gensymb}
\usepackage{tensor}
\usepackage[dvips,final]{graphicx}

\DeclareMathOperator{\e}{e} 
\DeclareMathOperator{\sgn}{sgn}

\newcommand{\dbar}{{\mathchar'26\mkern-11mu\mathrm{d}}}

\renewcommand{\geq}{\geqslant}

\newcommand{\ds}{\mathrm{d}}

\newcommand{\ms}{\mathrm{m}}

\newcommand{\rs}{\mathrm{r}}

\newcommand{\Es}{\mathrm{E}}

\newcommand{\Js}{\mathrm{J}}

\newcommand{\Lcal}{\ensuremath{\mathcal{L}}}

\newcommand{\Ocal}{\ensuremath{\mathcal{O}}}

\newcommand{\intfb}[1]{\ensuremath{\int\dbar^4k \,}}
\newcommand{\inttb}[1]{\ensuremath{\int\dbar^3k \,}}

\newcommand{\pdbd}[2]{\ensuremath{\frac{\partial #1}{\partial #2}}}

\newlength\rringshift 
\newcommand*\mathrring[1]{%
   \setbox0=\hbox{$#1$}%
   \dimen0 \wd0
   \advance\dimen0 -\rringshift
   \wd0 \dimen0
   \mathring{\copy0}%
   \kern-\wd0
   \kern \rringshift
   \mathring{%
     \phantom{\copy0}%
   }%
}

\begin{document}
\title{Stability of Cosmological Solution in $f(R)$ Models of Gravity}
\author{Ignacy Sawicki$^{1,2}$ and Wayne Hu$^{1}$}
\email{sawickii@theory.uchicago.edu} \affiliation{
{}$^1$ Kavli Institute for Cosmological Physics, Department of Astronomy \& Astrophysics, Enrico Fermi
Institute,  University of Chicago, Chicago IL 60637 \\
{}$^2$ Department of Physics,  University of Chicago, Chicago IL 60637}
\date{\today}

\begin{abstract}
    We reconcile seemingly conflicting statements in the literature about the
 behavior of cosmological solutions in modified theories of gravity where the Einstein-Hilbert Lagrangian for gravity is modified by the addition of a function of the Ricci scalar,  $f(R)$.
Using the example of $f(R) = \pm\mu^4/R$ we show that only such choices of $f(R)$ where $\ds^2f/\ds R^2>0$ have stable high-curvature
     limits and well-behaved cosmological solutions with a proper era of matter domination. The remaining models enter a phase dominated by both matter and scalar kinetic energy where the scalar curvature remains low.
\end{abstract}

\maketitle

The addition of a new term $f(R)$ to the Einstein-Hilbert action provides a possible explanation
for the late-time acceleration of the universe \cite{Caretal03,CapCarTro03,NojOdi03}.
However there is some confusion in the literature as to what kind of functions $f(R)$ result in
 solutions that are stable and cosmologically viable. In \cite{Song:2006ej}, the authors claim that a cosmology which behaves like $\Uplambda$CDM at early times and is stable to perturbations
 requires $f_{RR} \equiv \ds^2f/\ds R^2 \geq 0$. On the other hand, Bean {\it et al.}~\cite{Bean:2006up} have demonstrated that linear perturbations are stable in the CDTT model \cite{Caretal03} for which $f_{RR} < 0$. In addition, Amendola {\it et al.}~\cite{AmePolTsu06a,AmePolTsu06b} have claimed that $f(R)$ models do not have a normal matter-dominated epoch.

The aim of this {\it Brief Report}
 is to clarify the situation. We show that the sign of $f_{RR}$ \emph{does} determine whether the theory approaches the GR limit at high curvatures. For $f_{RR} > 0$, the theory at high curvatures behaves very close to GR and is stable. For $f_{RR}<0$, the GR limit is
 unstable.  In an $f(R)$ model,
 high density does not necessarily correspond to high curvature.  We will show that
 the Ricci scalar rapidly evolves to a low curvature solution at high redshift
 in this case. This is the instability discovered in \cite{Song:2006ej}. However, the low-curvature solution itself is stable, and hence it has well-behaved linear perturbations \cite{Bean:2006up}. The background cosmology in this low-curvature phase does not have the usual period of matter domination with $H\propto a^{-3/2}$. Instead it enters
 a  $\phi$MDE phase \cite{AmePolTsu06a,AmePolTsu06b} where $H \propto a^{-2}$, and is not observationally
 viable.

\medskip
{\it Preliminaries ---} We consider modifications to the Einstein-Hilbert action of the form \cite{Sta80}
\begin{equation}
    S=\int \ds^4x \sqrt{-g^\Js}\left[ \frac{R_\Js+f(R_\Js)}{2 \kappa^2}+{\cal L}_{\rm m} \right]\,,
\end{equation}
where $R_\Js$ is the Ricci scalar, $\kappa^2 = 8\pi G$ and $\Lcal_\ms$ is the matter Lagrangian. We define $f_R \equiv \ds f/\ds R_\Js$ and $f_{RR} \equiv \ds^2 f/\ds R_\Js^2$. ``J'' denotes the
Jordan frame, to be distinguished from the Einstein frame below.
We consider only the metric-formalism version of the theory, in which the Ricci scalar is a fully dynamical degree of freedom.

In order for an $f(R_\Js)$ model to be consistent with cosmological observations, the evolution of the scale factor $a_{\Js}$
must remain close to that of the concordance $\Uplambda$CDM model: i.e.\ a period of radiation domination, followed by high-curvature
matter domination, ending in an era of acceleration. In \cite{Song:2006ej}, the authors showed
that there exists an $f(R_\Js)$ that reproduces any desired expansion history.  The function is
parameterized by
\begin{equation}
    B(a_\Js) \equiv \frac{f_{RR}}{1+f_{R}} \frac{ \ds R_\Js }{\ds \ln a_\Js} \left(  \frac{ \ds \ln H_\Js}{\ds \ln a_\Js} \right)^{-1}\, ,
\end{equation}
where $H_\Js$ is the Hubble parameter.
However, models with $B<0$ were found to be unstable to linear perturbations at high curvature and hence to not reproduce the desired phenomenology.

We compare this finding with that in the Einstein frame in which stable cosmological
solutions for $B<0$ models have been found \cite{Bean:2006up}.
As emphasized by \cite{Magnano:1993bd,Chi03}, in the Einstein frame
gravity is unmodified but there is an additional minimally coupled scalar which evolves in a potential. Provided that $f_{RR} \neq 0$ and $1+f_R>0$, we can use the conformal transformation to the Einstein metric, $g^\Es_{\mu\nu} = \exp(\sigma\psi)g_{\mu\nu}^\Js$ with the new scalar field defined as
\begin{equation}
    \psi \equiv \sigma\ln (1+f_R) \label{e:psi} \, ,
\end{equation}
where $\sigma = \sgn(f_{R})$ ensures that $\psi \geq 0$. The Jordan-frame curvature $R_\Js(\psi)$ can then be found by inverting the relation \eqref{e:psi}. The resulting Einstein-frame
action is one for a scalar-tensor theory of gravity \footnote{Defining $\varphi \equiv \sigma\psi/\sqrt{2/3} \kappa$ yields canonical normalization.}
\begin{align}
    S &=\int\ds^4x \sqrt{-g^\Es}\Bigg\{\frac{1}{2\kappa^2}  \left[  R_\Es - \frac{3}{2}(\nabla_\Es\psi)^2 - V(\psi) \right]    \notag\\ & \quad+\exp(-2\sigma\psi)\Lcal_\ms[g^\Es_{\mu\nu}\exp(-\sigma\psi)] \Bigg\} \,,
\end{align}
where
\begin{equation}
    V(\psi(R_\Js)) = \frac{R_\Js f_R-f}{(1+f_R)^2} \,. \label{e:ef}
\end{equation}
``E'' denotes variables in the Einstein frame.
Here the matter is not minimally coupled, which results in the need to transform back to the Jordan frame for the interpretation of physical results.   Henceforth all derivatives, with the exception of $f_{R}$ and $f_{RR}$,  will be taken with respect to Einstein-frame variables.

\medskip
{\it High curvature stability ---}
For expository purposes we use a generalization of the CDTT model \cite{Caretal03}:
\begin{equation}
    f(R_\Js) = -\sigma \frac{\mu^4}{R_\Js} \,,
\end{equation}
where $\sigma = \pm 1$ and $\mu \sim H_0$. The $\sigma = +1$ model is the original CDTT choice  with $B<0$ and has a period of late-time acceleration with an effective $w=-2/3$. As in all inverse curvature models, the
behavior in the low curvature acceleration phase where $|f_R| \gg 1$ is independent of $\sigma$
\cite{Mena:2005ta}.
However  the acceleration phase for $\sigma=-1$ (mCDTT) eventually causes a sign change in
$1+f_R$ evolving {\it past} a coordinate singularity in the Einstein frame.  Moreover,
a sign change in $1+f_R$ and hence $B$ at low curvature
 would lead to Jordan-frame instabilities in a perturbed universe.
Nevertheless the mCDTT model is illustrative and its pathologies are avoidable.
Any function $f(R)$ such that
$
    \lim_{R_\Js\rightarrow \infty} f_R = 0
$
could have been chosen for this analysis. For example, the model $f(R) = \mu^2(R/\mu^2)^m$ with $0<m<1$ has $B>0$ and exhibits stable and observationally viable acceleration \cite{Faulkner:2006ub}.

We analyze the model in the Einstein frame, with the scalar defined as in \eqref{e:psi}.
At high curvature, $\psi$ is inversely related to the Jordan-frame Ricci scalar $R_\Js$
\begin{equation}
    \lim_{R_\Js \rightarrow \infty} \psi \sim \frac{\mu^4}{(R_\Js)^2} \,.
\end{equation}
Small $\psi$  implies a high curvature. If $\psi$ evolves away from zero, the expansion
enters a low-curvature phase. The potential for the scalar, as defined in \eqref{e:ef} is
\begin{equation}
    V(\psi) = 2\sigma \mu^2 \e^{-2\sigma\psi}\sqrt{\sigma(\e^{\sigma\psi}-1)} \,. \label{e:v}
\end{equation}

We assume that the matter is a sum of a pressureless component,
 $p_{\ms}=0$,  with density $\rho_\ms$ and an ultrarelativistic component, $p_{\rs} = 1/3$, with density $\rho_\rs$. Non-minimal coupling to the trace of the energy-momentum tensor
 requires that the evolution of $\psi$ is driven by an effective potential
\begin{equation}
    \Box_\Es\psi = \frac{1}{3} \pdbd{V}{\psi} - \frac{\sigma}{3}\kappa^2\rho_\ms^\Js \e^{-2\sigma\psi} \equiv \pdbd{V_\text{eff}}{\psi} \label{e:bp} \,,
\end{equation}
where we have used $\rho_\ms^\Js = \rho_\ms^\Es\e^{2\sigma\psi}$ to
bring out the dependence on $\psi$. Neither of the two quantities is conserved in the Robertson-Walker background of the Einstein frame. For example, $\rho_\ms^\Es$ evolves according to
\begin{equation}
    {\rho^\Es_\ms}' +3\rho_\ms^\Es = -\frac{\sigma}{2}\psi'\rho_\ms^\Es\,. \label{e:cons}
\end{equation}

\begin{figure}[ht]\begin{centering}
\includegraphics[width=\columnwidth]{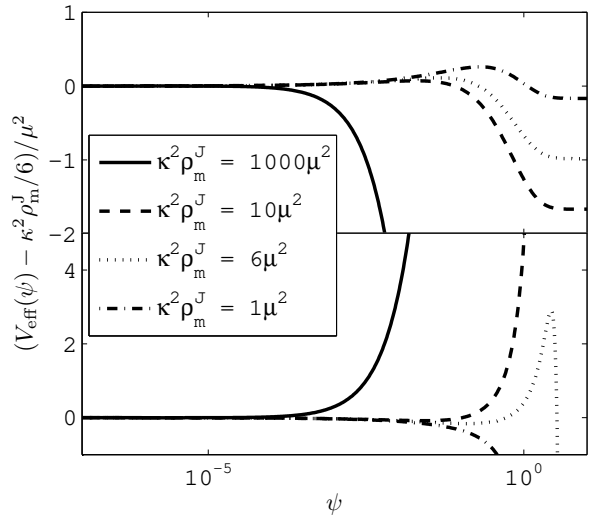}
\caption{Exact effective potentials with constant term subtracted for a range of matter densities
 for scalar field in CDTT (top panel) and mCDTT (bottom panel) models. Both models exhibit extrema at $R_\Js = \kappa^2\rho_\ms^\Js$: a maximum for CDTT and a minimum for mCDTT. \label{f:pots}}
\end{centering}\end{figure}

Assuming that $\psi \ll 1$, i.e.\ that the curvature is high---$R_\Js \gg \mu^2$---we can
expand the effective potential around $\psi =0$:
\begin{equation}
    V_\text{eff} \approx \frac{\kappa^2\rho_\ms^\Js}{6} + \frac{2}{3}\sigma\mu^2\sqrt{\psi} - \frac{1}{3}\sigma\kappa^2\rho_\ms^\Js\psi + \Ocal(\psi^{3/2})\,.
\end{equation}
This potential has an extremum at $\psi_{\rm GR} \approx \mu^4/(\kappa^2\rho_\ms^\Js)^{2}$, i.e.\ at the GR limit of $R_\Js \approx \kappa^2\rho_\ms^\Js$. For the CDTT model this is a maximum, whereas for the mCDTT model this is a minimum. The potentials for both models are presented in Fig.~\ref{f:pots}.

\begin{figure}[ht]\begin{centering}
\includegraphics[width=\columnwidth]{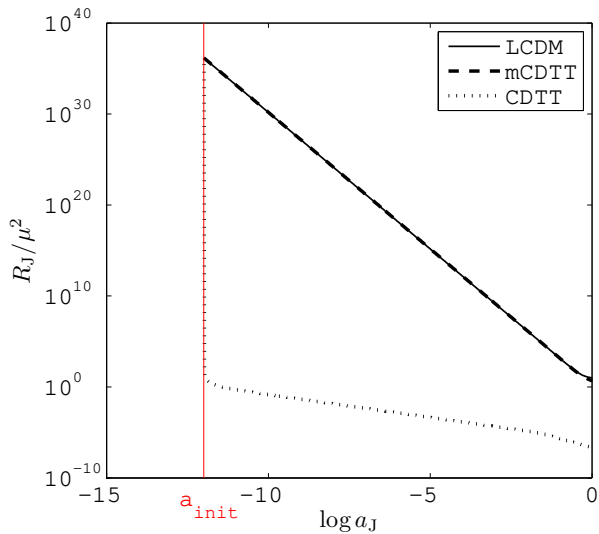}

\caption{Evolution of the Jordan-frame Ricci scalar for LCDM (solid; $\mu = H_0$),
mCDTT (dashed; $\alpha=2$) and CDTT (dotted; $\alpha=2$). All models are started with the same GR initial conditions. This solution is unstable in the CDTT model and the curvature rapidly relaxes to the $\phi$MDE phase where it is much lower than for the corresponding GR solution. This phase replaces the standard matter domination for $f(R)$ models with $B<0$. mCDTT follows standard LCDM evolution until late times, where it asymptotes to a milder
acceleration phase. \label{f:RJ}}
\end{centering}\end{figure}

We can easily demonstrate the instability of the high-curvature solution for the CDTT model in the Einstein frame. For $\psi \ll 1$,
the equation of motion of the scalar field becomes \eqref{e:bp}
\begin{equation}
    \psi'' + \left(3+\frac{H_\Es'}{H_\Es}\right) \psi'  + \frac{\sigma}{3H_\Es^2}\left( \frac{\mu^2}{\sqrt\psi}-\kappa^2\rho_\ms^\Js\right) \approx 0 \,, \label{e:pevol}
\end{equation}
where $' \equiv \ds /\ds \ln a_\Es$, and $H_\Es$
obeys the Friedmann equation
\begin{equation}
    H_\Es^2 = H_\Es^2 \frac{\psi'^2}{4}
    +
     \frac{1}{6}V(\psi) + \frac{\kappa^2\rho_\ms^\Es}{3} + \frac{\kappa^2\rho_\rs^\Es}{3} \,. \label{e:hub}
\end{equation}
We now assume that there exists a well behaved high-curvature solution to \eqref{e:pevol}, say $\psi_0$. We then perturb around this solution by defining
   $ \psi = \psi_0  + \delta\psi $
which results in the linearized equation of motion for the perturbation
\begin{equation}
    \delta\psi'' + \left( 3 + \frac{H'_\Es}{H_\Es}\right) \delta \psi' - \frac{\sigma\mu^2}{6H_\Es^2 \psi_0^{3/2}}\delta\psi = 0 \,.
\end{equation}
For $\sigma = +1$, $\delta\psi$ is unstable on very short timescales since
$\psi_{0} \ll 1$.
However, for the mCDTT model the high-curvature solution is stable, provided it exists. This analysis is related to \cite{Chi03} where the mass of the scalar around $\psi =0$
\begin{equation}
    m_\psi^2 =V_{,\psi\psi}|_{\psi=0} = f_{RR}^{-1} -3R -4f\,,
\end{equation}
gave a large tachyonic value $m_\psi^2 < 0$ for the CDTT model at high curvature.
 The field  rapidly rolls away from this maximum to high values, causing the Jordan frame curvature to be significantly below that of GR. We illustrate this
with a numerical solution in Fig.~\ref{f:RJ}. The calculation is initialized at the GR value of the Ricci scalar but it drops quickly to a small value and proceeds to decay as $a_\Js^{-4}$.

We now need to establish the existence of the high-curvature solution in the case of mCDTT, i.e.\ that corrections to GR remain small throughout the matter-dominated phase. We continue to assume that $\psi, \psi' \ll 1$ (implying that Jordan- and Einstein-frame quantities are approximately equal) and we  ignore the contribution of radiation.
We take $\psi_\mathrm{GR} = \mu^4/(\kappa^2\rho_\ms^\Js)^2$ as the GR solution for the Jordan-frame Ricci scalar.
Defining the perturbation away from this solution as $\delta\psi \equiv \psi - \psi_\mathrm{GR}$, we obtain the linearized equation for these perturbations from \eqref{e:pevol} during matter domination
\begin{equation}
\delta\psi'' + \frac{3}{2}\delta\psi' + \frac{9}{2\alpha^4 a_\Es^6}\delta\psi = -45\psi_\text{GR} \approx -\frac{45\alpha^4 a_\Es^6}{9} \,,
\end{equation}
where $\alpha^2 \equiv 3\mu^2/ \kappa^2 \rho_\ms^\Js(a_J=1)$.
This is a damped harmonic oscillator with an extremely high frequency, $\omega \sim 1/\alpha^2 a_\Es^3$. The amplitude is driven by the inhomogeneous term, giving
\begin{equation}
    \frac{\delta\psi}{\psi_\text{GR}} = - \frac{10\alpha^4}{ a_{\Es}^6}\,.  
\end{equation}
This remains extremely small throughout matter domination, confirming that for the mCDTT model the cosmology remains extremely close to that for GR until the acceleration phase.

\medskip
{\it Low-curvature $\phi$MDE phase --- }
We have demonstrated above that in the $B>0$ mCDTT model the high-curvature solution is stable, whereas it is not in the $B<0$ CDTT model. The evolution of $f(R)$ theories generically exhibits a fixed point in which the Einstein-frame scalar field is kinetic-energy dominated and remains a constant fraction of the energy density, together with the matter \cite{AmePolTsu06a,AmePolTsu06b}.
This phase replaces the ordinary matter dominated expansion for $B<0$ and is called the
$\phi$MDE solution. The analysis was extended in \cite{Amendola:2006we} to show that $B>0$ models do have usual matter-dominated eras for at least some initial conditions.

In this section we will demonstrate---using a slight modification of the method originally introduced in \cite{Copeland:1997et} and then used in \cite{Bean:2006up}---that indeed the $\phi$MDE phase is a saddle point in the evolution for both CDTT and mCDTT, but it can only be achieved for CDTT.

We introduce the dimensionless variables
\begin{equation}
x = \frac{\psi'}{2}\,, \qquad y = \frac{V}{6H_\Es^2} \,. \\
\end{equation}
Assuming that radiation is negligible, we can rewrite the Friedman equation \eqref{e:hub} as an equation for the matter energy density and its derivative
\begin{align}
    \frac{\kappa^2\rho_\ms^\Es}{3H^2_\Es} = 1 - x^2 - y\,,  \quad
    \frac{H'_\Es}{H_\Es} = -\frac{3}{2}(1+x^2 - y) \,. \label{e:Hp}
\end{align}
We assume that the potential for the scalar field is of exponential form, $V(\psi) = A\exp(\gamma\psi)$, which is valid when $\psi \gg 1$. From \eqref{e:v},
 $\gamma = -3/2$ for CDTT, while $\gamma = 2$ for mCDTT. This allows us to recast \eqref{e:bp} and \eqref{e:cons} as evolution equations for $x$ and $y$:
\begin{align}
    x' &= \frac{\sigma}{2} - \frac{3}{2}x - \frac{\sigma}{2}x^2 + \frac{3}{2}x^3 - y\left( \gamma + \frac{\sigma}{2} + \frac{3}{2}x \right)\,, \\
    y' &= y\left(2\gamma x + 3(1+ x^2 - y )\right) \,.
\end{align}
The fixed points of the evolution occur wherever $x' = y' = 0$. In particular, the point corresponding to $\phi$MDE lies at $(x,y) = (\sigma/3, 0)$. At this point, the scalar field energy density is completely dominated by the kinetic term and the fraction of the total energy in matter is $\Omega_\ms = 8/9$. From equation \eqref{e:Hp}, the effective equation of state is $w = 1/9$, resulting in
the Einstein-frame Hubble parameter evolving as $H_\Es \propto a_\Es^{-5/3}$. The Jordan and
Einstein frame Hubble parameter are related by
\begin{equation}
    H_\Js = H_\Es \e^{\sigma\psi/2} \left(1-\frac{\sigma\psi'}{2} \right) \,.
\end{equation}
Since $\psi \sim 2\ln a_\Es/3$ at this fixed point, $H_\Js \propto a_\Js^{-2}$, just as during radiation domination.

Using \eqref{e:pevol} and \eqref{e:hub}, we can express the Jordan-frame Ricci scalar as
\begin{equation}
    R_\Js = \e^{\sigma\psi}\left(4V + 2\sigma\pdbd{V}{\psi}\right)\,,
\end{equation}
and hence both the models under consideration have $R_\Js \ll H^2$ at the $\phi$MDE fixed point.

The eigenvalues at this point are $-4/3$ and $2/3(5+\alpha\sigma)$. For both models $\phi$MDE is a saddle point. However, only for the CDTT model is this stage of evolution possible to achieve. Here, the field has a large value and is running off to positive infinity, exponentially suppressing the contribution of the potential to the Hubble parameter.
In the mCDTT model, $V\propto \e^{2\psi}$: the large-field limit is also the large-potential limit. The existence of the $\phi$MDE phase therefore requires that $\psi \ll 0$, which is not a valid limit of the Jordan-frame theory and is a result of the use of the exponential approximation to the potential.

\medskip
{\it Discussion --- }
We have shown that in the case of the generalized CDTT model, the sign of $B$ determines the behavior of the model and leads to a qualitative change in the Einstein-frame potential for the scalar field. This scalar field has a large negative mass squared in the high curvature
 limit for $B<0$ models, resulting in unstable solutions in the GR limit. The field runs off to a large value, where the potential is effectively exponential. The cosmology enters a phase of simultaneous scalar field and matter domination with the physical Hubble parameter, $H_\Js$, proportional to $a_\Js^{-2}$ and a very small value of the curvature scalar in the Jordan frame.

The situation is radically different in the case of mCDTT, where $B>0$ at high curvature.
The GR-like solution is stable, the cosmology goes through a phase of standard matter domination where
corrections to GR remain small.  The acceleration behavior is asymptotically the same as
CDTT  but this specific model will develop instabilities deep in the acceleration regime.

Another important difference between the two model classes parameterized by $B$ is their behavior around mass sources, i.e.\ their predictions for Solar-System tests. Indeed, in $B<0$ models the Ricci scalar is perturbed only slightly from its low
 background value by sources of matter density, $R_\Js \ll \kappa^2\rho_\ms$.  PPN parameters can then be calculated using the weak-field approximations \cite{EriSmiKam06, Chiba:2006jp} and which results in a highly excluded value of $\gamma = 1/2$ \cite{Chi03}. However, this is not necessarily true for $B>0$ models. For these models, the value of the Ricci scalar can be high, reflecting the local matter density. It is therefore inappropriate to linearize the perturbations to $R_\Js$. As shown in \cite{Faulkner:2006ub} and \cite{Navarro:2006mw}, the chameleon mechanism resulting from the non-minimal coupling of the theory to matter and the existence of a minimum in the scalar potential around the GR value of the Ricci allows for the theory to evade Solar-System tests, at least for certain functions $f(R)$ and certain ranges of parameters. We will elaborate in detail on the constraints put on $f(R)$ theories  by solar-system tests of gravity in a future work \cite{Hu:2007ax}.

\medskip
{\it Acknowledgments ---} We thank R. Bean, S. Carroll, H. Peiris, J. Santiago, Y.S. Song, J. Weller and A. Upadhye for useful conversations.  This
work was supported by  the KICP under NSF PHY-0114422, DOE under
DE-FG02-90ER-40560, and the
David and Lucile Packard Foundation.

\hfill
\bibliography{fr_stab}

\begin{thebibliography}{18}
\expandafter\ifx\csname natexlab\endcsname\relax\def\natexlab#1{#1}\fi
\expandafter\ifx\csname bibnamefont\endcsname\relax
  \def\bibnamefont#1{#1}\fi
\expandafter\ifx\csname bibfnamefont\endcsname\relax
  \def\bibfnamefont#1{#1}\fi
\expandafter\ifx\csname citenamefont\endcsname\relax
  \def\citenamefont#1{#1}\fi
\expandafter\ifx\csname url\endcsname\relax
  \def\url#1{\texttt{#1}}\fi
\expandafter\ifx\csname urlprefix\endcsname\relax\def\urlprefix{URL }\fi
\providecommand{\bibinfo}[2]{#2}
\providecommand{\eprint}[2][]{\url{#2}}

\bibitem[{\citenamefont{Carroll et~al.}(2004)\citenamefont{Carroll, Duvvuri,
  Trodden, and Turner}}]{Caretal03}
\bibinfo{author}{\bibfnamefont{S.~M.} \bibnamefont{Carroll}},
  \bibinfo{author}{\bibfnamefont{V.}~\bibnamefont{Duvvuri}},
  \bibinfo{author}{\bibfnamefont{M.}~\bibnamefont{Trodden}}, \bibnamefont{and}
  \bibinfo{author}{\bibfnamefont{M.~S.} \bibnamefont{Turner}},
  \bibinfo{journal}{Phys. Rev.} \textbf{\bibinfo{volume}{D70}},
  \bibinfo{pages}{043528} (\bibinfo{year}{2004}), \eprint{astro-ph/0306438}.

\bibitem[{\citenamefont{Capozziello et~al.}(2003)\citenamefont{Capozziello,
  Carloni, and Troisi}}]{CapCarTro03}
\bibinfo{author}{\bibfnamefont{S.}~\bibnamefont{Capozziello}},
  \bibinfo{author}{\bibfnamefont{S.}~\bibnamefont{Carloni}}, \bibnamefont{and}
  \bibinfo{author}{\bibfnamefont{A.}~\bibnamefont{Troisi}}
  (\bibinfo{year}{2003}), \eprint{astro-ph/0303041}.

\bibitem[{\citenamefont{Nojiri and Odintsov}(2003)}]{NojOdi03}
\bibinfo{author}{\bibfnamefont{S.}~\bibnamefont{Nojiri}} \bibnamefont{and}
  \bibinfo{author}{\bibfnamefont{S.~D.} \bibnamefont{Odintsov}},
  \bibinfo{journal}{Phys. Rev.} \textbf{\bibinfo{volume}{D68}},
  \bibinfo{pages}{123512} (\bibinfo{year}{2003}), \eprint{hep-th/0307288}.

\bibitem[{\citenamefont{Song et~al.}(2007)\citenamefont{Song, Hu, and
  Sawicki}}]{Song:2006ej}
\bibinfo{author}{\bibfnamefont{Y.-S.} \bibnamefont{Song}},
  \bibinfo{author}{\bibfnamefont{W.}~\bibnamefont{Hu}}, \bibnamefont{and}
  \bibinfo{author}{\bibfnamefont{I.}~\bibnamefont{Sawicki}},
  \bibinfo{journal}{Phys. Rev.} \textbf{\bibinfo{volume}{D75}},
  \bibinfo{pages}{044004} (\bibinfo{year}{2007}), \eprint{astro-ph/0610532}.

\bibitem[{\citenamefont{Bean et~al.}(2007)\citenamefont{Bean, Bernat, Pogosian,
  Silvestri, and Trodden}}]{Bean:2006up}
\bibinfo{author}{\bibfnamefont{R.}~\bibnamefont{Bean}},
  \bibinfo{author}{\bibfnamefont{D.}~\bibnamefont{Bernat}},
  \bibinfo{author}{\bibfnamefont{L.}~\bibnamefont{Pogosian}},
  \bibinfo{author}{\bibfnamefont{A.}~\bibnamefont{Silvestri}},
  \bibnamefont{and} \bibinfo{author}{\bibfnamefont{M.}~\bibnamefont{Trodden}},
  \bibinfo{journal}{Phys. Rev.} \textbf{\bibinfo{volume}{D75}},
  \bibinfo{pages}{064020} (\bibinfo{year}{2007}), \eprint{astro-ph/0611321}.

\bibitem[{\citenamefont{Amendola
  et~al.}(2007{\natexlab{a}})\citenamefont{Amendola, Polarski, and
  Tsujikawa}}]{AmePolTsu06a}
\bibinfo{author}{\bibfnamefont{L.}~\bibnamefont{Amendola}},
  \bibinfo{author}{\bibfnamefont{D.}~\bibnamefont{Polarski}}, \bibnamefont{and}
  \bibinfo{author}{\bibfnamefont{S.}~\bibnamefont{Tsujikawa}},
  \bibinfo{journal}{Phys. Rev. Lett.} \textbf{\bibinfo{volume}{98}},
  \bibinfo{pages}{131302} (\bibinfo{year}{2007}{\natexlab{a}}),
  \eprint{astro-ph/0603703}.

\bibitem[{\citenamefont{Amendola et~al.}(2006)\citenamefont{Amendola, Polarski,
  and Tsujikawa}}]{AmePolTsu06b}
\bibinfo{author}{\bibfnamefont{L.}~\bibnamefont{Amendola}},
  \bibinfo{author}{\bibfnamefont{D.}~\bibnamefont{Polarski}}, \bibnamefont{and}
  \bibinfo{author}{\bibfnamefont{S.}~\bibnamefont{Tsujikawa}}
  (\bibinfo{year}{2006}), \eprint{astro-ph/0605384}.

\bibitem[{\citenamefont{Starobinsky}(1980)}]{Sta80}
\bibinfo{author}{\bibfnamefont{A.~A.} \bibnamefont{Starobinsky}},
  \bibinfo{journal}{Phys. Lett.} \textbf{\bibinfo{volume}{B91}},
  \bibinfo{pages}{99} (\bibinfo{year}{1980}).

\bibitem[{\citenamefont{Magnano and Sokolowski}(1994)}]{Magnano:1993bd}
\bibinfo{author}{\bibfnamefont{G.}~\bibnamefont{Magnano}} \bibnamefont{and}
  \bibinfo{author}{\bibfnamefont{L.~M.} \bibnamefont{Sokolowski}},
  \bibinfo{journal}{Phys. Rev.} \textbf{\bibinfo{volume}{D50}},
  \bibinfo{pages}{5039} (\bibinfo{year}{1994}), \eprint{gr-qc/9312008}.

\bibitem[{\citenamefont{Chiba}(2003)}]{Chi03}
\bibinfo{author}{\bibfnamefont{T.}~\bibnamefont{Chiba}},
  \bibinfo{journal}{Phys. Lett.} \textbf{\bibinfo{volume}{B575}},
  \bibinfo{pages}{1} (\bibinfo{year}{2003}), \eprint{astro-ph/0307338}.

\bibitem[{\citenamefont{Mena et~al.}(2006)\citenamefont{Mena, Santiago, and
  Weller}}]{Mena:2005ta}
\bibinfo{author}{\bibfnamefont{O.}~\bibnamefont{Mena}},
  \bibinfo{author}{\bibfnamefont{J.}~\bibnamefont{Santiago}}, \bibnamefont{and}
  \bibinfo{author}{\bibfnamefont{J.}~\bibnamefont{Weller}},
  \bibinfo{journal}{Phys. Rev. Lett.} \textbf{\bibinfo{volume}{96}},
  \bibinfo{pages}{041103} (\bibinfo{year}{2006}), \eprint{astro-ph/0510453}.

\bibitem[{\citenamefont{Faulkner et~al.}(2006)\citenamefont{Faulkner, Tegmark,
  Bunn, and Mao}}]{Faulkner:2006ub}
\bibinfo{author}{\bibfnamefont{T.}~\bibnamefont{Faulkner}},
  \bibinfo{author}{\bibfnamefont{M.}~\bibnamefont{Tegmark}},
  \bibinfo{author}{\bibfnamefont{E.~F.} \bibnamefont{Bunn}}, \bibnamefont{and}
  \bibinfo{author}{\bibfnamefont{Y.}~\bibnamefont{Mao}} (\bibinfo{year}{2006}),
  \eprint{astro-ph/0612569}.

\bibitem[{\citenamefont{Amendola
  et~al.}(2007{\natexlab{b}})\citenamefont{Amendola, Gannouji, Polarski, and
  Tsujikawa}}]{Amendola:2006we}
\bibinfo{author}{\bibfnamefont{L.}~\bibnamefont{Amendola}},
  \bibinfo{author}{\bibfnamefont{R.}~\bibnamefont{Gannouji}},
  \bibinfo{author}{\bibfnamefont{D.}~\bibnamefont{Polarski}}, \bibnamefont{and}
  \bibinfo{author}{\bibfnamefont{S.}~\bibnamefont{Tsujikawa}},
  \bibinfo{journal}{Phys. Rev.} \textbf{\bibinfo{volume}{D75}},
  \bibinfo{pages}{083504} (\bibinfo{year}{2007}{\natexlab{b}}),
  \eprint{gr-qc/0612180}.

\bibitem[{\citenamefont{Copeland et~al.}(1998)\citenamefont{Copeland, Liddle,
  and Wands}}]{Copeland:1997et}
\bibinfo{author}{\bibfnamefont{E.~J.} \bibnamefont{Copeland}},
  \bibinfo{author}{\bibfnamefont{A.~R.} \bibnamefont{Liddle}},
  \bibnamefont{and} \bibinfo{author}{\bibfnamefont{D.}~\bibnamefont{Wands}},
  \bibinfo{journal}{Phys. Rev.} \textbf{\bibinfo{volume}{D57}},
  \bibinfo{pages}{4686} (\bibinfo{year}{1998}), \eprint{gr-qc/9711068}.

\bibitem[{\citenamefont{Erickcek et~al.}(2006)\citenamefont{Erickcek, Smith,
  and Kamionkowski}}]{EriSmiKam06}
\bibinfo{author}{\bibfnamefont{A.~L.} \bibnamefont{Erickcek}},
  \bibinfo{author}{\bibfnamefont{T.~L.} \bibnamefont{Smith}}, \bibnamefont{and}
  \bibinfo{author}{\bibfnamefont{M.}~\bibnamefont{Kamionkowski}},
  \bibinfo{journal}{Phys. Rev.} \textbf{\bibinfo{volume}{D74}},
  \bibinfo{pages}{121501} (\bibinfo{year}{2006}), \eprint{astro-ph/0610483}.

\bibitem[{\citenamefont{Chiba et~al.}(2006)\citenamefont{Chiba, Smith, and
  Erickcek}}]{Chiba:2006jp}
\bibinfo{author}{\bibfnamefont{T.}~\bibnamefont{Chiba}},
  \bibinfo{author}{\bibfnamefont{T.~L.} \bibnamefont{Smith}}, \bibnamefont{and}
  \bibinfo{author}{\bibfnamefont{A.~L.} \bibnamefont{Erickcek}}
  (\bibinfo{year}{2006}), \eprint{astro-ph/0611867}.

\bibitem[{\citenamefont{Navarro and Van~Acoleyen}(2006)}]{Navarro:2006mw}
\bibinfo{author}{\bibfnamefont{I.}~\bibnamefont{Navarro}} \bibnamefont{and}
  \bibinfo{author}{\bibfnamefont{K.}~\bibnamefont{Van~Acoleyen}}
  (\bibinfo{year}{2006}), \eprint{gr-qc/0611127}.

\bibitem[{\citenamefont{Hu and Sawicki}(2007)}]{Hu:2007ax}
\bibinfo{author}{\bibfnamefont{W.}~\bibnamefont{Hu}} \bibnamefont{and}
  \bibinfo{author}{\bibfnamefont{I.}~\bibnamefont{Sawicki}},
  \bibinfo{journal}{\prd} \textbf{\bibinfo{volume}{\rm submitted}}
  (\bibinfo{year}{2007}), \eprint{arXiv:0705.1158 [astro-ph]}.

\end{thebibliography}
\end{document}